\documentclass[a4paper,11pt]{article}
\pdfoutput=1 

\usepackage{jheppub} 

\usepackage[T1]{fontenc}
\usepackage{multirow,bbold,slashed,wasysym}
\usepackage[greek,english]{babel}

\allowdisplaybreaks
 
\definecolor{nicered}{rgb}{0.7,0.1,0.1} 
\definecolor{nicegreen}{rgb}{0.1,0.5,0.1}
\definecolor{niceblue}{rgb}{0.0,0.1,0.7}
\hypersetup{colorlinks,citecolor=niceblue,linkcolor=niceblue,urlcolor=niceblue}

\usepackage[normalem]{ulem}

\def \bm#1{\mbox{\boldmath$#1$\unboldmath}}
\def \beq{\begin{equation}}
\def \eeq{\end{equation}}
\def \bea{\begin{eqnarray}}
\def \eea{\end{eqnarray}}

\title{How large are hadronic contributions to $\bm{h \to \gamma \gamma}$?}

\author[a]{Ulrich Haisch}

\affiliation[a]{Max Planck Institute for Physics, \\ Boltzmannstr.~8, 85748 Garching, Germany}

\emailAdd{haisch@mpp.mpg.de}

\preprint{MPP-2025-190} 

\abstract{The decay of the Higgs boson into two photons, $h \to \gamma \gamma$, is a loop-induced process within the Standard Model, predominantly mediated by loops of $W$ bosons and top quarks. While these leading contributions are well understood, the role of hadronic effects, which arise from non-perturbative QCD dynamics, has received less attention, with recent studies reporting puzzling and contradictory results. In this work, we present a systematic evaluation of the hadronic contributions to the $h \to \gamma \gamma$ decay width using dispersion relations. Our analysis shows that these contributions are exceedingly small, as expected, altering the decay width by about $0.004\%$ under conservative assumptions. Therefore, hadronic effects can be safely neglected even in the context of future high-precision Higgs measurements at current and next-generation colliders. As an aside, we also estimate the possible size of hadronic contributions to Higgs production in gluon-gluon~fusion.}

\begin{document} 
\maketitle
\flushbottom

\section{Introduction} 
\label{sec:introduction}

The decay of the Higgs boson into two photons, $h \to \gamma \gamma$, was one of the key discovery channels at the Large Hadron Collider (LHC)~\cite{ATLAS:2012yve,CMS:2012qbp} and remains central to precision studies of its properties. Future measurements at the high-luminosity LHC (HL-LHC) are expected to determine the branching ratio for the $h \to \gamma \gamma$ channel with a precision at the percent level~\cite{ATL-PHYS-PUB-2019-006,ATL-PHYS-PUB-2022-018,ATL-PHYS-PUB-2025-014,ATL-PHYS-PUB-2025-018}. To fully exploit such measurements, equally precise Standard Model~(SM) predictions are required. In fact, achieving such a theoretical precision came at the price of no small amount of theoretical blood, sweat, and tears~\cite{Ellis:1975ap,Shifman:1979eb,Zheng:1990qa,Djouadi:1990aj,Dawson:1992cy,Melnikov:1993tj,Inoue:1994jq,Spira:1995rr,Steinhauser:1996wy,Fleischer:2004vb,Fugel:2004ug,Degrassi:2005mc,Harlander:2005rq,Aglietti:2006tp,Passarino:2007fp,Maierhofer:2012vv,Sturm:2014nva,Niggetiedt:2020sbf,Davies:2021zbx,Boito:2022fmn,Sang:2025jfl}. The state-of-the-art SM prediction reported in~\cite{Davies:2021zbx} incorporates perturbative QCD corrections up to four loops and electroweak~(EW) corrections up to two loops, resulting in a combined theoretical uncertainty of about $1.7\%$ for the $h \to \gamma \gamma$ decay width. The residual uncertainty is dominated by the missing next-next-to-leading order~(NNLO) EW and mixed QCD-EW corrections~\cite{Sang:2025jfl}, which would need to be computed to reduce the total uncertainty to $1\%$ or below.

The calculations~\cite{Ellis:1975ap,Shifman:1979eb,Zheng:1990qa,Djouadi:1990aj,Dawson:1992cy,Melnikov:1993tj,Inoue:1994jq,Spira:1995rr,Steinhauser:1996wy,Fleischer:2004vb,Fugel:2004ug,Degrassi:2005mc,Harlander:2005rq,Aglietti:2006tp,Passarino:2007fp,Maierhofer:2012vv,Sturm:2014nva,Niggetiedt:2020sbf,Davies:2021zbx,Boito:2022fmn,Sang:2025jfl} have all considered perturbative contributions to the $h \to \gamma \gamma$ decay width arising from Feynman loop diagrams, with the dominant effects coming from loop momenta of order the SM Higgs-boson mass $125 \, \text{GeV}$. For the light quarks, such perturbative effects are strongly suppressed by two powers of their masses, rendering them phenomenologically irrelevant. However, at very low energies of order $1 \, \text{GeV}$, the effects of confinement and strong interactions between quarks become important. As a result, light quarks can no longer be treated as free particles in theoretical calculations, because their interactions are dominated by complex, non-perturbative QCD effects. Light-quark loops may therefore also give rise to a non-perturbative contribution to the $h \to \gamma \gamma$ decay width. The potential magnitude of such effects has recently been studied in~\cite{Hernandez-Juarez:2025ees,Knecht:2025nyo}. While~\cite{Hernandez-Juarez:2025ees} found the non-perturbative light-quark contributions to vanish,~\cite{Knecht:2025nyo} reported a~$3\%$ correction to the $h \to \gamma \gamma$ decay width. At face value, both results appear inconsistent with the expected quadratic mass suppression of light-quark contributions, which predicts a perturbative effect of roughly $0.002\%$, primarily due to strange-quark loops. 

Motivated by the puzzling findings of~\cite{Hernandez-Juarez:2025ees,Knecht:2025nyo}, this article offers an independent assessment of non-perturbative QCD contributions to the $h \to \gamma \gamma$ decay width. Our analysis employs a dispersive framework consistent with the symmetries of low-energy QCD, analyticity, and unitarity, expressing the hadronic contributions as a convolution of light-quark scalar and energy-momentum tensor form factors of the intermediate hadronic states~$X$, weighted by the $S$-wave cross sections for $X \to \gamma \gamma$. By combining theoretical input with experimental data, this method enables a systematic extraction of the hadronic effects. Our~approach shares important features with the dispersive method for evaluating the leading hadronic contribution to the muon anomalous magnetic moment~$\big($$a_\mu^{\text{had}, \text{LO}}$$\big)$, which relies on $e^+ e^- \to \text{hadrons}$ data as well as the analyticity and unitarity of the vacuum polarization function of the photon. For~recent analyses of this type, see, for instance,~\cite{Davier:2017zfy,Keshavarzi:2018mgv,Colangelo:2018mtw,Hoferichter:2019mqg,Davier:2019can,Keshavarzi:2019abf,Aoyama:2020ynm}. A~concise introduction to dispersive methods, along with their contemporary applications to low-energy phenomenology within the SM, is provided, for example, in the nice, brand-new lecture notes~\cite{Colangelo:2025sah}.

The structure of this work is as follows: Section~\ref{sec:preliminaries} reviews the SM Higgs effective interactions at low energies and the perturbative contributions to the $h \to \gamma \gamma$ amplitude. Section~\ref{sec:non-perturbative_effects} introduces the theoretical framework for computing the corresponding hadronic effects. Section~\ref{sec:phenomenology} details the ingredients of our prediction and briefly discusses both the limitations of the method and the main sources of uncertainty. Finally, Section~\ref{sec:conclusions} summarizes our main findings along with providing a brief outlook. An alternative, although less accurate, method for estimating the hadronic contributions to the $h \to \gamma \gamma$ decay width is discussed in Appendix~\ref{app:LMD}. While we’re at it, Appendix~\ref{app:ggF} finally throws in a similar estimate of the corresponding hadronic effects in Higgs production via gluon-gluon fusion~(ggF). Without further ado, let's crack straight~into~it!

\section{Preliminaries} 
\label{sec:preliminaries}

To set the stage, we examine the effective interactions of the SM Higgs field $h$ for energies below the charm-quark threshold. The corresponding Lagrangian, describing its couplings to photons, light quarks ($u$, $d$, and $s$), and gluons, is given by
\beq \label{eq:Lagrangian}
{\cal L} = \left [ C_{\gamma \gamma} F_{\mu \nu} F^{\mu \nu} - \frac{7}{9} \left ( Q_\Gamma + Q_\Delta \right ) - \frac{2}{9} \hspace{0.5mm} Q_\theta \right ] \frac{h}{v} \,, 
\eeq
where 
\beq \label{eq:Qi}
Q_\Gamma = m_u \bar u u + m_d \bar d d \,, \qquad Q_\Delta = m_s \bar s s \,, \qquad Q_\theta = -\frac{9 \alpha_s}{8 \pi} \hspace{0.5mm} G_{\mu \nu}^a G^{a, \mu \nu} + Q_\Gamma + Q_\Delta \,, 
\eeq
denotes, respectively, the combined scalar current of the up and down quarks, the scalarcurrent of the strange quark, and the trace of the energy-momentum tensor. The latter originates from the QCD conformal anomaly~\cite{Crewther:1972kn,Chanowitz:1972vd,Shifman:1979eb}. Above, $F_{\mu \nu}$ and $G_{\mu \nu}^a$ denote the electromagnetic and QCD field strength tensors, respectively, while $\alpha_s$ is the strong-coupling constant, $v$ the Higgs vacuum expectation value, and $m_q$~the light quark masses. Note~that the effective Higgs coupling to gluons in~(\ref{eq:Qi}) arises from integrating out the heavy quarks~($t$, $b$, and~$c$). 

\begin{table}[t!]
\centering
\begin{tabular}{|c|c|c|c|}
\hline
Parameter & Value & Parameter & Value \\ \hline
$\alpha$ & $1/137.036$ & $m_c$ & $1.273 \, \text{GeV}$ \\
$v$ & $246.22 \, \text{GeV}$ & $m_b$ & $4.183 \, \text{GeV}$ \\
$m_u$ & $2.16 \, \text{MeV}$ & $m_\pi$ & $140 \, \text{MeV}$ \\
$m_d$ & $4.70 \, \text{MeV}$ & $m_K$ & $496 \, \text{MeV}$ \\
$m_s$ & $93.5 \, \text{MeV}$ & $m_h$ & $125.2 \, \text{GeV}$ \\
 \hline
\end{tabular}
\vspace{2mm} 
\caption{The set of input parameters used in this analysis. All values are taken from the most recent Particle Data Group~(PDG)~review~\cite{ParticleDataGroup:2024cfk}. The light-quark masses ($m_u$, $m_d$, and $m_s$) are quoted in the $\overline{\text{MS}}$ scheme at a scale of $2 \, \text{GeV}$, whereas the charm- and bottom-quark masses ($m_c$ and $m_b$) are given in the $\overline{\text{MS}}$ scheme evaluated at their own mass scales.}
\label{tab:input_parameters}
\end{table}

The Wilson coefficient $C_{\gamma \gamma}$ in~(\ref{eq:Lagrangian}) captures the perturbative SM contributions to the effective Higgs-photon coupling. It is connected to the $h \to \gamma \gamma$ decay width via
\beq \label{eq:width}
\Gamma ( h \to \gamma \gamma ) = \frac{m_h^3}{4 \pi v^2} \, \big | C_{\gamma \gamma} \big |^2 = 9.284 \left (1 \pm 1.69\% \right ) \, \text{keV} \,, 
\eeq
where $m_h$ is the Higgs-boson mass, and the numerical value on the right-hand side is taken from~\cite{Davies:2021zbx}. With the value of 
$m_h$ from Table~\ref{tab:input_parameters}, and neglecting the quoted uncertainty in~(\ref{eq:width}), this yields
\beq \label{eq:Cgammagamma}
C_{\gamma \gamma} \simeq -1.90 \cdot 10^{-3} \,, 
\eeq
where the sign is fixed by an explicit one-loop calculation. In fact, in the limit of an infinitely heavy $W$~boson and top quark, the dominant contributions to the Wilson coefficient above takes the approximate form 
\beq \label{eq:Wboson_topquark}
C_{\gamma \gamma}^W \simeq -\frac{\alpha}{8 \pi} \hspace{0.75mm} 7 \,, \qquad 
C_{\gamma \gamma}^t \simeq \frac{\alpha}{8 \pi} \hspace{0.5mm} \frac{16}{9} \,,
\eeq
where $\alpha$ is the electromagnetic coupling constant. In contrast, the contributions of the bottom, charm, and light quarks can be approximated as
\beq \label{eq:bottom_charm_lightquarks}
C_{\gamma \gamma}^q \simeq \frac{\alpha}{8 \pi} \left \{ 6 \hspace{0.25mm} e_q^2 \hspace{0.5mm} \frac{m_q^2}{m_h^2} \hspace{0.5mm} \left [ \ln^2 \left ( -\frac{m_q^2}{m_h^2} \right ) - 4 \right ] \right \} \,,
\eeq
with $e_q$ the quark electric charge, $e_u = 2/3$ for up-type and $e_d = -1/3$ for down-type quarks. A few remarks regarding~(\ref{eq:Wboson_topquark}) and~(\ref{eq:bottom_charm_lightquarks}) are in order. First, $W$-boson loops provide the dominant perturbative contribution to the $h \to \gamma \gamma$ decay amplitude in the SM, while the top-quark contribution is smaller by roughly a factor of four and interferes destructively. Second, the perturbative contributions of the bottom, charm, and light quarks are strongly mass-suppressed, proportional to $m_q^2/m_h^2$. This scaling occurs because the triangle diagram involving two vector currents and a scalar insertion proportional to $m_q$ vanishes unless there is an extra chirality flip, which in perturbation theory is provided by the quark mass.

To estimate the perturbative contributions from the bottom, charm, and light quarks, we define
\beq \label{eq:delta_q}
\delta_q = \frac{\left | C_{\gamma \gamma}^q - C_{\gamma \gamma} \right |^2}{\left | C_{\gamma \gamma} \right |^2} - 1 \,, 
\eeq
and, using the input parameters from Table~\ref{tab:input_parameters}, we obtain
\beq \label{eq:deltas}
\delta_b \simeq 0.74\% \,, \quad \delta_c \simeq 0.59\% \,, \quad \delta_s \simeq 0.0022\% \,, \quad \delta_d \simeq 1.2 \cdot 10^{-7} \,, \quad \delta_u \simeq 1.1 \cdot 10^{-7} \,.
\eeq
This shows that the combined perturbative contributions of the bottom and charm quarks are at the percent level, whereas the contributions from all light quarks are entirely negligible for phenomenological purposes, at only a few tenths of a part per million. Consequently, light-quark contributions are omitted both in the state-of-the-art prediction for the $h \to \gamma \gamma$ decay width reported in~\cite{Davies:2021zbx} and in~(\ref{eq:Cgammagamma}).

\section{Non-perturbative effects} 
\label{sec:non-perturbative_effects}

The perturbative contributions to the $h \to \gamma \gamma$ decay width, discussed in~(\ref{eq:width}) to~(\ref{eq:deltas}), originate from the standard triangle diagrams, with the dominant effects coming from loop momenta of order $m_h$. At scales below the QCD confinement scale, $\Lambda_{\text{QCD}} = \mathcal{O}(1 \, \text{GeV})$, non-perturbative effects become relevant. The Higgs can first couple to light-quark pairs, which hadronize into mesons such as pions or kaons before decaying into two photons. By summing over all such intermediate hadronic states, $X = \pi^+ \pi^-$, $\pi^0 \pi^0$, $\pi^0 \eta$, $K^+ K^-$, $K_S K_S$, $\ldots \hspace{0.5mm}$, and accounting for their strong interactions, one can estimate the size of the non-perturbative contributions. Realize that the same framework can be applied to the calculation of $a_\mu^{\text{had}, \text{LO}}$, which also arises from non-perturbative QCD effects associated with light-quark loops. Via a dispersion relation, these contributions are related to the cross section for~$e^+ e^- \to \text{hadrons}$, effectively summing over all intermediate hadronic states. For~recent precision extractions of $a_\mu^{\text{had}, \text{LO}}$, see, for instance,~\cite{Davier:2017zfy,Keshavarzi:2018mgv,Colangelo:2018mtw,Hoferichter:2019mqg,Davier:2019can,Keshavarzi:2019abf,Aoyama:2020ynm}. Some of the techniques used in the analysis of $a_\mu^{\text{had}, \text{LO}}$ therefore also apply to the computation of non-perturbative effects in the $h \to \gamma \gamma$ decay width, as outlined in the following. 

Let $C_{\gamma \gamma, i}^{\text{had}} (s)$ denote the Wilson coefficient encoding the hadronic contributions of the operators in~(\ref{eq:Qi}) to the $h \to \gamma \gamma$ decay amplitude. Since it is an analytic function of $s$ with a branch cut beginning at the two-pion threshold, $s = 4 m_\pi^2$, where $m_\pi$~is the pion mass, this Wilson coefficient naturally admits a dispersion relation of the form
\beq \label{eq:dispersion_relation}
C_{\gamma \gamma, i}^{\text{had}} (m_h^2) = \frac{1}{\pi} \int_{4 m_\pi^2}^\infty \! ds \, \frac{{\text{Im}} \, C_{\gamma \gamma, i}^{\text{had}} (s)}{s - m_h^2 - i \epsilon} \,,
\eeq
with $\epsilon > 0$ and infinitesimally small. The imaginary part in the numerator originates from on-shell intermediate states $X$, which connect the vacuum to the $\gamma \gamma$ final state via the operator $Q_i$. By invoking unitarity, one can thus write
\beq \label{eq:unitarity}
\text{Im} \, C_{\gamma \gamma, i}^{\text{had}} (s) = \frac{1}{2} \sum_{X = \text{hadrons}} \int \! d \Phi_X \hspace{0.25mm} F_i^X (s) \hspace{0.25mm} T^\ast_{X \to \gamma \gamma} (s) \,, 
\eeq
where $d \Phi_X$ denotes the phase-space measure for the intermediate state $X$, and 
\begin{gather} 
F_i^X (s) = \big | F_i^X (s) \big | \hspace{0.5mm} e^{i \phi_i^X (s)} \equiv \left \langle 0 \left | Q_i \right | X \right \rangle \,, \label{eq:FiX} \\[2mm]
T_{X \to \gamma \gamma} (s) = \big | T_{X \to \gamma \gamma} (s) \big | \hspace{0.5mm} e^{i \tau_X (s)} \equiv \left \langle X | \gamma \gamma \right \rangle \,, \label{eq:TX} 
\end{gather}
represent the corresponding form factor and transition amplitude, respectively. Both the form factors and the transition amplitudes are complex quantities, with their corresponding strong phases denoted by $\phi_i^X (s)$ and $\tau_X (s)$. 

\begin{figure}[t!]
\centering
\includegraphics[width=0.95\textwidth]{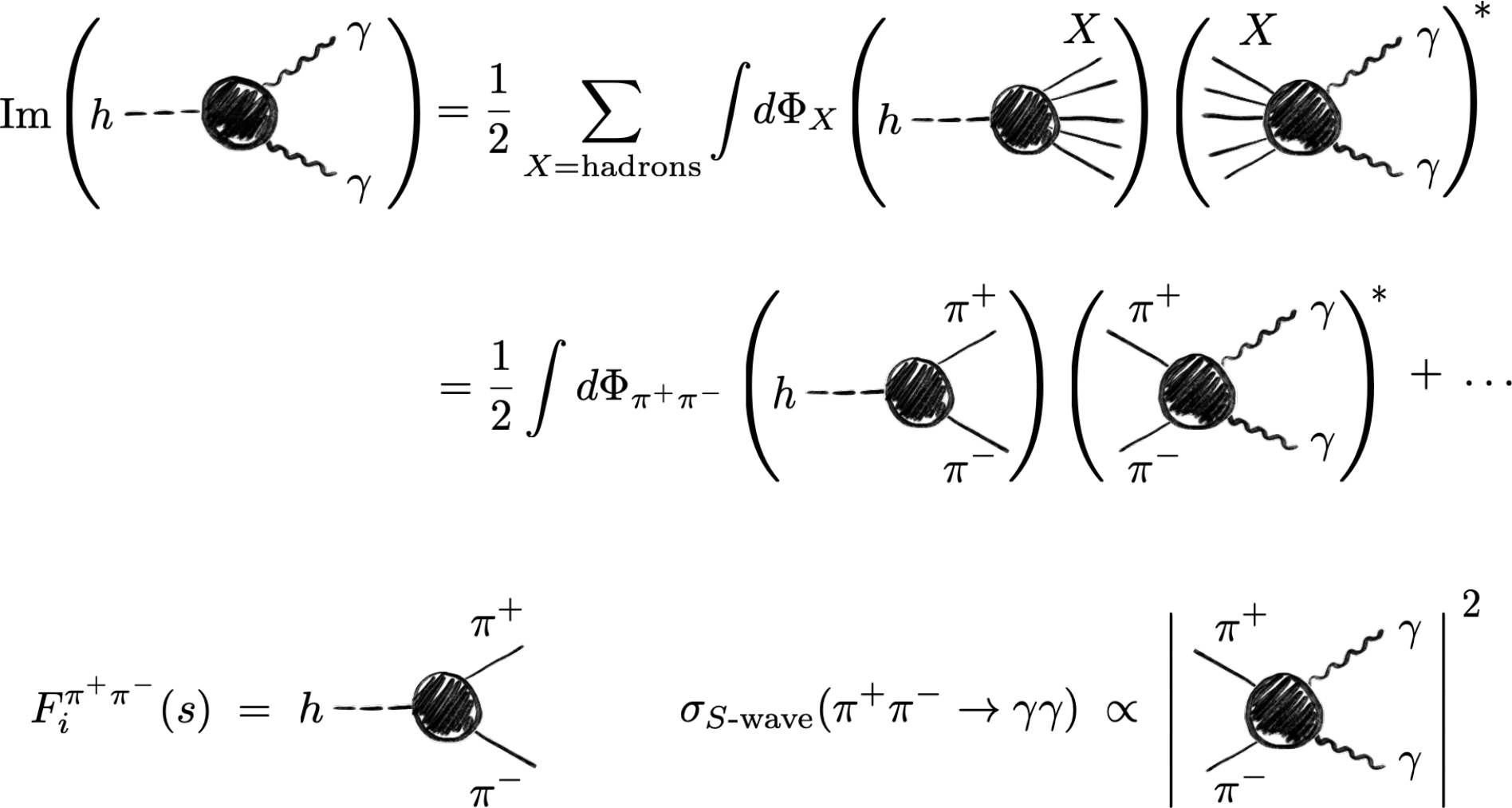}
\vspace{2mm} 
\caption{Graphical illustration of the optical theorem applied to the hadronic contribution of the $h \to \gamma \gamma$ amplitude, as given in~(\ref{eq:unitarity}). The ellipses in the second line denote all hadronic intermediate states $X$ other than $\pi^+ \pi^-$. Also shown are diagrammatic representations of the form factors $F_i^{\pi^+ \pi^-} (s)$ and the $S$-wave cross section $\sigma_{\text{$S$-wave}} \left ( \pi^+ \pi^- \to \gamma \gamma \right )$. \label{fig:illustration}}
\end{figure}

The dominant contributions to the sum over $X$ in~(\ref{eq:unitarity}) generally arise from hadronic two-body intermediate states, such as $\pi^+ \pi^-$. In this case 
\beq \label{eq:dPhiX}
 d \Phi_{\pi^+ \pi^-} = \frac{\beta_\pi (s)}{16 \pi^2} \, d \Omega \,, \qquad \beta_{\pi} (s) = \sqrt{1 - \frac{4 m_\pi^2}{s}} \,, 
\eeq
where $\beta_{\pi} (s)$ represents the velocity of each charged pion in the center-of-mass frame and $d \Omega = d \cos \theta \hspace{0.25mm} d\phi$ is the solid-angle element. Since the operators $Q_i$ have spin $0$ and positive parity, like the SM Higgs boson, they couple to hadronic states in an $S$-wave configuration. The~transition amplitude $T_{\pi^+ \pi^- \to \gamma \gamma} (s)$ can therefore be expressed in terms of the corresponding $S$-wave cross section as
\beq \label{eq:transition_amplitude}
T_{\pi^+ \pi^- \to \gamma \gamma} (s) = \sqrt{\frac{2 \pi \beta_\pi (s)}{s} \hspace{0.5mm} \sigma_{\text{$S$-wave}} \left (\pi^+ \pi^- \to \gamma \gamma \right )} \; e^{i \tau_{\pi^+ \pi^-} (s)} \,. 
\eeq 
Analogous formulas apply to all other hadronic two-body intermediate states. For identical particles, however, an additional symmetry factor of $1/2$ must be included in both the phase space and the $S$-wave cross section. Also note that, since the $S$-wave cross section does not depend on the polar angle $\theta$ or the azimuthal angle $\phi$, the integral over $d \Omega$ factorizes and simply evaluates to $4 \pi$. Figure~\ref{fig:illustration} provides a graphical illustration of the relations and definitions given in~(\ref{eq:unitarity}),~(\ref{eq:FiX}),~(\ref{eq:TX}), and~(\ref{eq:transition_amplitude}).

Based on the preceding discussion, the result in~(\ref{eq:unitarity}) can be approximated as
\beq \label{eq:unitarity_approximation}
\text{Im} \, C_{\gamma \gamma, i}^{\text{had}} (s) \simeq \sum_{X = \text{two-body}} \kappa_i^X (s) \hspace{0.5mm} e^{i \left ( \phi_i^X (s) - \tau_X (s) \right )} \,, 
\eeq
where the sum over $X$ runs over all hadronic two-body intermediate states, and for later convenience, we have defined:
\beq \label{eq:kappaiX}
\kappa_i^X (s) \equiv \big | F_i^X (s) \big | \hspace{0.5mm} \sqrt{\frac{S_X \beta_X^3 (s)}{8 \pi s} \hspace{0.5mm} \sigma_{\text{$S$-wave}} \left (X \to \gamma \gamma \right )} \,.
\eeq
Here, $\beta_X(s)$ is defined analogously to~(\ref{eq:dPhiX}), and $S_X$ denotes the symmetry factor of the hadronic two-body intermediate state. For example, $S_{\pi^+ \pi^-} = 1$ while $S_{\pi^0 \pi^0} = 1/2$. Three- and higher-body hadronic intermediated are further phase-space suppressed and hence expected to provide a subleading effect to $\text{Im} \, C_{\gamma \gamma, i}^{\text{had}} (s)$. 

The Wilson coefficient $C_{\gamma \gamma, i}^{\text{had}} (s)$ in~(\ref{eq:dispersion_relation}) is, in general, a complex quantity. For our purposes, however, it is sufficient to consider only its magnitude. Assuming maximal constructive interference among all intermediate two-body hadronic states, we can neglect the phase factors in~(\ref{eq:unitarity_approximation}) to obtain the approximation
\begin{gather} 
\big | C_{\gamma \gamma, i}^{\text{had}} (m_h^2) \big | \simeq \!\! \sum_{X = \text{two-body}} \! \big | C_{\gamma \gamma, i}^{\text{had}, X} (m_h^2) \big | \,, \label{eq:unitarity_approximation_bound1} \\[2mm]
\big | C_{\gamma \gamma, i}^{\text{had}, X} (m_h^2) \big | = \frac{1}{\pi} \int_{4 m_\pi^2}^\infty \! ds \, \frac{\kappa_i^X (s)}{\left | s - m_h^2 - i \epsilon \right |} \,, \label{eq:unitarity_approximation_bound2}
\end{gather}
which, in fact, provides an upper bound on the exact contribution from all intermediate two-body hadronic states. Note that in practice the integral over $s$ must be cut off at some point to avoid including the perturbative effects in~(\ref{eq:bottom_charm_lightquarks}). In our numerical analysis presented in~Section~\ref{sec:phenomenology}, we impose a cut-off at $s = 4 \, \text{GeV}^2$, since the contribution from charm-quark loops is treated as purely perturbative.

In terms of the contributions in~(\ref{eq:unitarity_approximation_bound1}), the magnitude of the full hadronic correction to the Wilson coefficient $C_{\gamma \gamma}$ appearing in~(\ref{eq:width}) can be written as
\beq \label{eq:Cgammagammahad}
\big | C_{\gamma \gamma}^{\text{had}} \big | = \frac{7}{9} \, \Big ( \big | C_{\gamma \gamma, \Gamma}^{\text{had}} (m_h^2) \big | + \big | C_{\gamma \gamma, \Delta}^{\text{had}} (m_h^2) \big | \Big ) + \frac{2}{9} \, \big | C_{\gamma \gamma, \theta}^{\text{had}} (m_h^2) \big | \,.
\eeq
Note that the numerical prefactors in~(\ref{eq:Cgammagammahad}) follow from~(\ref{eq:Lagrangian}), where they multiply the operators $Q_\Gamma$, $Q_\Delta$, and $Q_\theta$ introduced in~(\ref{eq:Qi}). In the next section, we will use~(\ref{eq:unitarity_approximation_bound1}), (\ref{eq:unitarity_approximation_bound2}), and~(\ref{eq:Cgammagammahad}) to estimate the magnitude of the hadronic contributions to the $h \to \gamma \gamma$ decay~width. 

\section{Phenomenological analysis} 
\label{sec:phenomenology}

The calculations of the form factors $F_i^X (s)$ has a checkered past --- see, for instance,~\cite{Voloshin:1985tc,Raby:1988qf,Truong:1989my,Donoghue:1990xh,Gunion:1989we,Monin:2018lee}. Modern calculations of the form factors, such as~\cite{Winkler:2018qyg,Blackstone:2024ouf}, are based on chiral perturbation theory~(ChPT) and extensively employ dispersion relations. This framework ensures consistency with the symmetries of low-energy QCD while systematically incorporating analyticity and unitarity, thereby enhancing the reliability of theoretical predictions over a broad kinematic range, extending from $s = 4 m_\pi^2$ up to around $s = 4 \, \text{GeV}^2$. In the following, we summarize the key results concerning the form factors that are relevant for this work.

In our numerical analysis, we consider the hadronic two-body intermediate states $X = \pi^+ \pi^-$, $\pi^0 \pi^0$, $\pi^0 \eta$, $K^+ K^-$, and $K_S K_S$. At leading order~(LO) in ChPT, the form factors of charged and neutral two-body intermediate states are approximately equal:
\beq \label{eq:Fpi_F_K}
\big | F_i^{\pi^+ \pi^-} (s) \big | \simeq \big | F_i^{\pi^0 \pi^0} (s) \big | \,, \qquad \big | F_i^{K^+ K^-} (s) \big | \simeq \big | F_i^{K_S K_S} (s) \big | \,.
\eeq
Higher-order corrections in ChPT introduce a mild energy dependence and small differences between charged and neutral form factors. For pions, these corrections scale as $s/(4 \pi f_\pi)^2$, where $f_\pi \simeq 130 \, \text{MeV}$ denotes the pion decay constant~\cite{ParticleDataGroup:2024cfk}. Analogous statements hold for the kaon case, with the appropriate substitutions. For the present analysis, we regard~(\ref{eq:Fpi_F_K}) as an adequate approximation. 

Due to isospin breaking, the scalar form factor for $\pi^0 \eta$ can be related to that of $\pi^0 \pi^0$, because the physical $\eta$ meson contains a small admixture of the neutral pion, described by the parameter
\beq \label{eq:epsilon}
\epsilon_{\pi^0\eta} \simeq \frac{\sqrt{3}}{4} \frac{m_d-m_u}{m_s - \hat m} \simeq 1.2\% \,, \qquad \hat m = \frac{1}{2} \left ( m_u + m_d \right ) \,, 
\eeq
whose numerical value has been obtained using the relevant input parameters in~Table~\ref{tab:input_parameters}. Consequently, the magnitudes of the form factors for the $\pi^0 \eta$ system can be approximated~as
\beq \label{eq:Fpieta}
\big | F_i^{\pi^0 \eta} (s) \big | \simeq \epsilon_{\pi^0\eta} \hspace{0.25mm} \big | F_i^{\pi^0 \pi^0} (s) \big | \,,
\eeq
with an accuracy that is more than sufficient for the present context.

\begin{figure}[t!]
\centering
\includegraphics[width=0.6\textwidth]{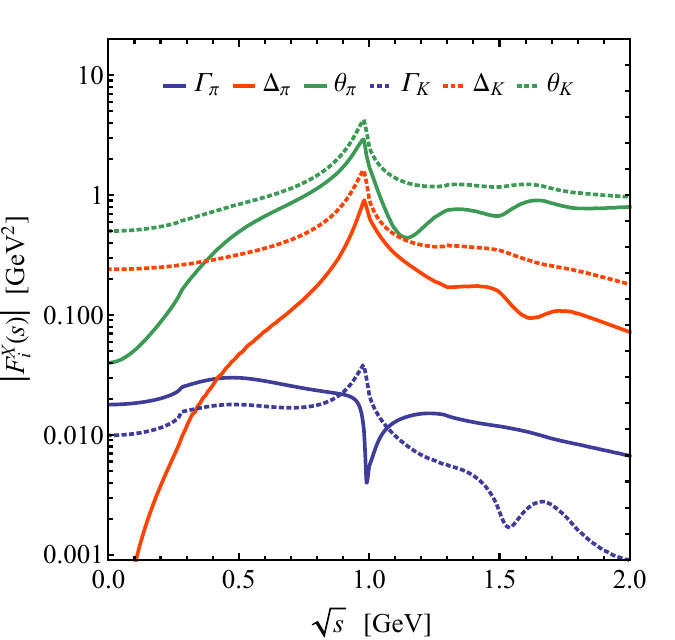}
\vspace{0mm} 
\caption{Magnitudes of the form factors $F_i^X(s)$ corresponding to the operators $Q_i$ introduced in~(\ref{eq:Qi}), shown for $X = \pi \pi, K K$. Further details are provided in the main text.} \label{fig:formfactors}
\end{figure}

Figure~\ref{fig:formfactors} displays the magnitudes of the form factors $F_i^X(s)$, associated with the operators $Q_i$ defined in~(\ref{eq:Qi}), for $X = \pi \pi, K K$, as obtained in~\cite{Winkler:2018qyg}. The results shown provide a reliable estimate of the form factors, exhibiting qualitative agreement with earlier calculations~\cite{Raby:1988qf,Truong:1989my,Donoghue:1990xh,Gunion:1989we,Monin:2018lee} and good agreement with the recent computation~\cite{Blackstone:2024ouf}. For more detailed comparisons, see~\cite{Winkler:2018qyg,Blackstone:2024ouf}. To put the numerical values for~$\big | F_i^X (s) \big |$, as read off from the plot, into perspective, we recall the LO ChPT results for all relevant form factors~\cite{Truong:1989my,Donoghue:1990xh}:
\begin{gather} \label{eq:LO_ChPT}
F^\pi_\Gamma (s) \simeq m_\pi^2 \,, \qquad F^\pi_\Delta (s) \simeq 0 \,, \qquad F^\pi_\theta (s) \simeq s + 2 m_\pi^2 \,, \\[2mm]
F^K_\Gamma (s) \simeq \frac{m_\pi^2}{2} \,, \qquad F^K_\Delta (s) \simeq m_K^2 - \frac{m_\pi^2}{2} \,, \qquad F^K_\theta (s) \simeq s + 2 m_K^2 \,.
\end{gather}
From a comparison, we find that the form-factor results shown in~Figure~\ref{fig:formfactors} are in rough agreement with the LO ChPT expressions in~(\ref{eq:LO_ChPT}) up to $\sqrt{s} \lesssim 0.5 \, \text{GeV}$. The only exception is $\big | F^\pi_\Delta (s) \big |$, which vanishes at LO in ChPT. This prevents a direct comparison with the plotted result, although a similar qualitative trend can still be observed in the figure. In~the region $\sqrt{s} \gtrsim 0.5 \, \text{GeV}$, the LO ChPT expressions begin to deviate from the results of~\cite{Winkler:2018qyg}, with the discrepancies becoming increasingly pronounced as $\sqrt{s} \simeq 1.0 \, \text{GeV}$. This is due to the strong influence of the $f_0 (980)$ resonance, which predominantly couples to the intermediate~$\pi \pi$ and $K K$ states~\cite{ParticleDataGroup:2024cfk}. This is physically reasonable, since the $f_0 (980)$ is a~spin-0 resonance with positive parity, isospin $0$, a mass of approximately $980 \, \text{MeV}$, and a peak width in~$\pi \pi$ of roughly~$50 \, \text{MeV}$. In fact, in~\cite{Winkler:2018qyg} the $f_0 (980)$ is treated using a full two-channel analysis that accounts for both elastic and inelastic $\pi \pi$ and $K K$ scatterings, while incorporating theoretical constraints in the form of Roy-Steiner equations~\cite{Roy:1971tc,Hite:1973pm}. The strong phases and the inelasticity parameter needed for such an analysis were extracted in~\cite{Winkler:2018qyg} from~\cite{Hoferichter:2012wf}. This provides reliable form factor values up to $\sqrt{s} \simeq 1.3 \, \text{GeV}$. For higher values of $\sqrt{s}$, the study~\cite{Winkler:2018qyg} employs the asymptotic conditions on the $T$-matrix derived in~\cite{Moussallam:1999aq}, which ensure that the $\pi \pi$ and $K K$ scattering phases exhibit the correct behavior in the limit~$\sqrt{s} \to \infty$. Note that for $\sqrt{s} \gtrsim 1.3 \, \text{GeV}$, the form factors obtained from the two-channel analysis become less reliable, since additional channels such as $4 \pi$ and $\eta \eta$ start to play a potentially important role in the computation of $F_i^X (s)$. 

From~Figure~\ref{fig:formfactors} it is furthermore evident that the magnitudes of the form factors exhibit a clear hierarchy: those arising from the operator $Q_\theta$ are largest, followed by $Q_\Delta$, and finally~$Q_\Gamma$. Numerically, for the $\pi \pi$ channel at $\sqrt{s} \simeq 1.0 \, \text{GeV}$, we obtain $1.7 \, \text{GeV}^2$, $0.7 \, \text{GeV}^2$, and $5 \cdot 10^{-3} \, \text{GeV}^2$, respectively. For $K K$ the corresponding values are $2.6 \, \text{GeV}^2$, $0.9 \, \text{GeV}^2$, and $2 \cdot 10^{-2} \, \text{GeV}^2$. From~(\ref{eq:unitarity_approximation_bound1}) and~(\ref{eq:unitarity_approximation_bound2}), it follows that hadronic contributions to the $h \to \gamma \gamma$ decay width are largely dominated by gluonic and strange-quark effects, while up- and down-quark contributions are close to negligible. We recall that the gluonic contributions, originating from the QCD conformal anomaly, are ultimately connected to the heavy-quark terms integrated~out~in~(\ref{eq:Lagrangian}).

\begin{figure}[t!]
\centering
\includegraphics[width=0.6\textwidth]{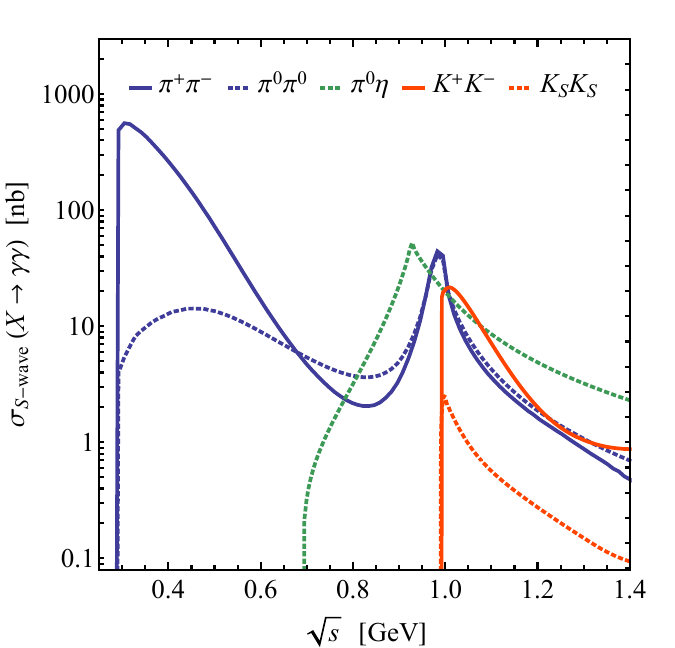}
\vspace{0mm} 
\caption{$S$-wave contributions to the $X \to \gamma \gamma$ cross sections for $X = \pi^+ \pi^-$, $\pi^0 \pi^0$, $\pi^0 \eta$, $K^+ K^-$, and $K_S K_S$. See the main text for further explanations.} \label{fig:crosssections}
\end{figure}

Significant theoretical~\cite{Bernabeu:2008wt,Oller:2008kf,Garcia-Martin:2010kyn,Moussallam:2011zg,Hoferichter:2011wk,Dai:2014lza,Dai:2014zta,Danilkin:2017lyn,Danilkin:2018qfn,Danilkin:2019opj,Lu:2020qeo,Danilkin:2020pak,Schafer:2023qtl,Stamen:2024ocm,Deineka:2024mzt} and experimental effort has also been devoted to determining the cross sections for the processes $X \to \gamma \gamma$. In this study, we rely on the determinations of the $S$-wave contributions obtained from the dispersion analyses of~\cite{Danilkin:2018qfn,Danilkin:2020pak,Deineka:2024mzt}, which make use of the $\gamma \gamma \to X$ data reported in~\cite{Dally:1980dj,Dally:1981ur,ARGUS:1989ird,CrystalBall:1990oiv,Boyer:1990vu,Belle:2009ylx,Belle:2013eck,Kuessner2022}. Figure~\ref{fig:crosssections} shows the $S$-wave contributions to the~$X \to \gamma \gamma$ cross sections for the relevant two-body hadronic intermediate states. A~couple of comments seem to be in order. First, for $\sqrt{s} \lesssim 0.8 \, \text{GeV}$, the $\pi \pi$ channels have the largest cross section, whereas for $\sqrt{s} \gtrsim 0.8 \, \text{GeV}$, this is generally no longer the case. Second, the~$S$-wave cross sections for the production of two charged mesons are always larger than those for their neutral counterparts. The~difference between charged and neutral channels can be easily understood in terms of tree-level couplings. In the charged channel, the process~$X \to \gamma \gamma$ proceeds at tree level via direct photon couplings, leading to a Born $S$-wave cross section of the form
\beq \label{eq:S-wave_Born}
\sigma_{\text{$S$-wave}}^{\text{Born}} \left (X \to \gamma \gamma \right ) = \frac{\pi \alpha^2 \beta_X (s)}{2 s} \left[ \frac{1 - \beta_X^2 (s)}{\beta_X (s)} \hspace{0.25mm} \ln \left( \frac{1 + \beta_X (s)}{1 - \beta_X (s)} \right) \right]^2 \,, 
\eeq
where $\beta_X(s)$ is defined similarly to~(\ref{eq:dPhiX}). In contrast, the neutral channel lacks a direct coupling to photons, so $X \to \gamma \gamma$ occurs only via higher-order processes such as charged meson loops or rescattering. Third, as in~Figure~\ref{fig:formfactors}, the $f_0 (980)$ resonance leaves a clear imprint on the~$\pi \pi$ and $K K$ cross sections, producing a characteristic distortion near $\sqrt{s} = 1 \, \text{GeV}$ due to its strong coupling to both the $\pi \pi$ and $K K$ channels. A similar effect is observed for the~$\pi^0 \eta \to \gamma \gamma$ process due to the $a_0(980)$ resonance, which, aside from having isospin $1$, has properties similar to those of the $f_0 (980)$~\cite{ParticleDataGroup:2024cfk}. 

\begin{figure}[t]
\centering
\includegraphics[width=0.6\textwidth]{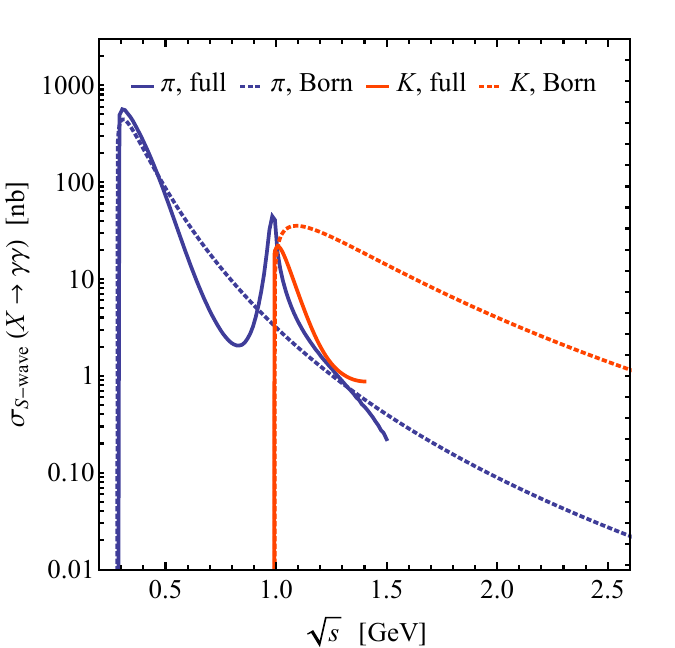}
\vspace{0mm} 
\caption{$S$-wave contributions to the $X \to \gamma \gamma$ cross sections for $X = \pi^+ \pi^-$ and $K^+ K^-$ are shown. The figure compares the full results from~\cite{Danilkin:2018qfn,Deineka:2024mzt} with the Born approximation given in~(\ref{eq:S-wave_Born}).} \label{fig:xsectionscomparison}
\end{figure}

\begin{figure}[t!]
\centering
\includegraphics[width=0.6\textwidth]{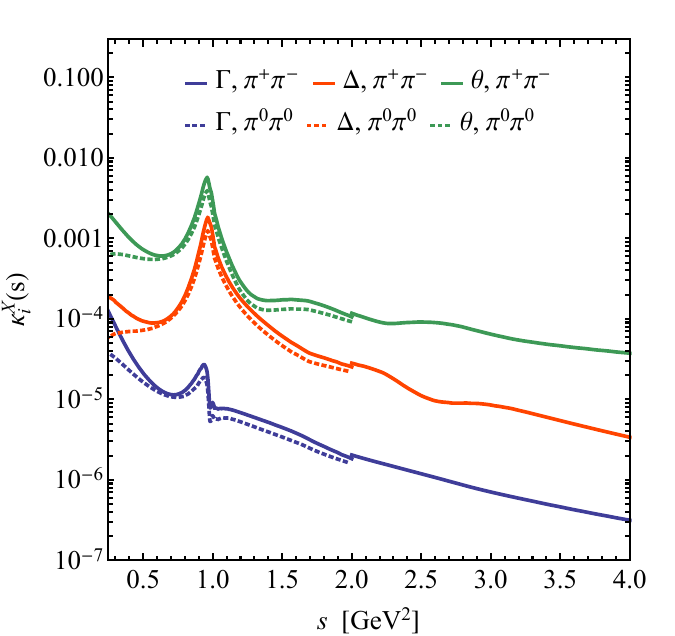} 

\vspace{0mm}

\includegraphics[width=0.6\textwidth]{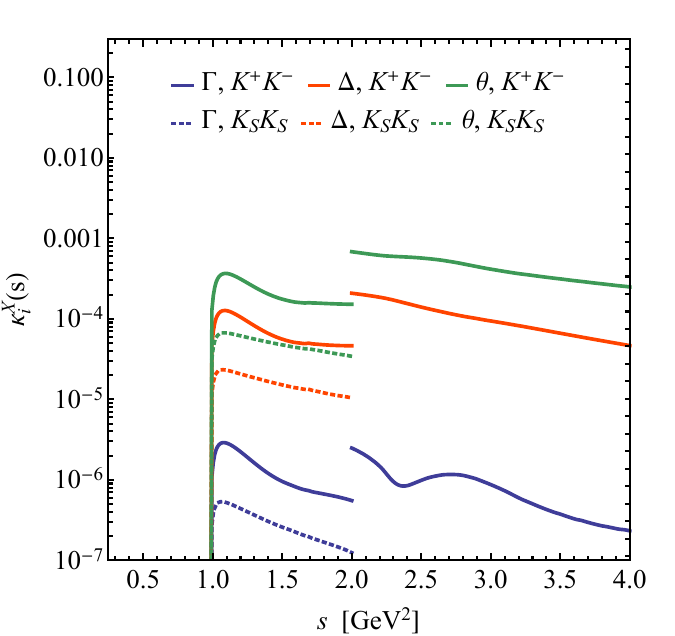}
\vspace{2mm} 
\caption{Functions $\kappa_i^X (s)$ introduced in~(\ref{eq:kappaiX}) for the operators $Q_i$ in~(\ref{eq:Qi}) and $X = \pi^+ \pi^-$, $\pi^0 \pi^0$, $K^+ K^-$, and $K_S K_S$. Up to $s = 2 \, \text{GeV}^2$, the results for charged pions and kaons are based on the full $S$-wave cross sections, whereas for $s > 2 \, \text{GeV}^2$ the Born approximation~(\ref{eq:S-wave_Born}) is used. Consult the main text for further details.} \label{fig:kappas}
\end{figure}

To gain a quantitative understanding of the significance of higher-order processes, such as charged meson loops or rescattering, we compare in~Figure~\ref{fig:xsectionscomparison} the full and Born results for the $S$-wave contributions to the $X \to \gamma \gamma$ cross sections for $X = \pi^+ \pi^-$ and $K^+ K^-$. For~the~$\pi^+ \pi^-$ case, the $f_0 (980)$ resonance produces a pronounced dip-peak structure, highlighting the role of interference effects in $\pi^+ \pi^- \to \gamma \gamma$ around $\sqrt{s} \simeq 1 \, \text{GeV}$. These effects arise from the interference between the leading Born amplitude and the loop-induced amplitudes that encode the strong $\pi^+ \pi^-$ and $K^+ K^-$ interactions. We also see that for~$K^+ K^-$, the two-kaon threshold introduces a sharp feature in the $K^+ K^- \to \gamma \gamma$ cross section near~$\sqrt{s} = 1 \, \text{GeV}$. Charged meson loops and rescattering dominate in this region, and lead to a pronounced destructive interference with the Born contribution above threshold. Finally, we emphasize that for $\sqrt{s} \gtrsim 1.4 \, \text{GeV}$ ($\sqrt{s} \gtrsim 1.0 \, \text{GeV}$) the Born cross sections consistently overestimate the full results for $\pi^+ \pi^-$ ($K^+ K^-$).

\begin{table}[t!]
\centering
\begin{tabular}{|c|c|c|c|}
\hline 
$4 m_\pi^2 < s < 2 \, \text{GeV}^2$ & $Q_\Gamma$ & $Q_\Delta$ & $Q_\theta$ \\
\hline
$\pi^+ \pi^-$ & $1.19 \cdot 10^{-9}$ & $7.98 \cdot 10^{-9}$ & $3.59 \cdot 10^{-8}$ \\
$\pi^0 \pi^0$ & $4.39 \cdot 10^{-10}$ & $5.45 \cdot 10^{-9}$ & $2.14 \cdot 10^{-8}$ \\
$\pi^0 \eta$ & $7.58 \cdot 10^{-12}$ & $9.42 \cdot 10^{-11}$ & $3.69 \cdot 10^{-10}$ \\
$K^+ K^-$ & $2.69 \cdot 10^{-11}$ & $1.43 \cdot 10^{-9}$ & $4.38 \cdot 10^{-9}$ \\
$K_S K_S$ & $6.00 \cdot 10^{-12}$ & $3.27 \cdot 10^{-10}$ & $1.00 \cdot 10^{-9}$ \\
\hline
$\Sigma$ & $1.67 \cdot 10^{-9}$ & $1.53 \cdot 10^{-8}$ & $6.31 \cdot 10^{-8}$ \\
\hline
\hline
$2 \, \text{GeV}^2 < s < 4 \, \text{GeV}^2$ & $Q_\Gamma$ & $Q_\Delta$ & $Q_\theta$ \\
\hline
$\pi^+ \pi^-$ & $3.50 \cdot 10^{-11}$ & $4.10 \cdot 10^{-10}$ & $2.75 \cdot 10^{-9}$ \\
$\pi^0 \pi^0$ & $2.47 \cdot 10^{-11}$ & $2.90 \cdot 10^{-10}$ & $1.94 \cdot 10^{-9}$ \\
$\pi^0 \eta$ & $4.27 \cdot 10^{-13}$ & $5.00 \cdot 10^{-12}$ & $3.36 \cdot 10^{-11}$ \\
$K^+ K^-$ & $3.39 \cdot 10^{-11}$ & $4.32 \cdot 10^{-9}$ & $1.77 \cdot 10^{-8}$ \\
$K_S K_S$ & $2.40 \cdot 10^{-11}$ & $3.06 \cdot 10^{-9}$ & $1.25 \cdot 10^{-8}$ \\
\hline
$\Sigma$ & $1.18 \cdot 10^{-10}$ & $8.08 \cdot 10^{-9}$ & $3.50 \cdot 10^{-8}$ \\
\hline
\end{tabular}
\vspace{2mm}
\caption{Values of the magnitudes $\big | C_{\gamma \gamma, i}^{\text{had}} (m_h^2) \big |$ defined in~(\ref{eq:unitarity_approximation_bound1}) and~(\ref{eq:unitarity_approximation_bound2}) for the operators in~(\ref{eq:Qi}) and the channels $X = \pi^+ \pi^-$, $\pi^0 \pi^0$, $\pi^0 \eta$, $K^+ K^-$, and $K_S K_S$. The rows labeled $\Sigma$ correspond to the sum over all channels for a given operator $Q_i$. The upper part of the table lists the values obtained using the full $S$-wave cross section in~(\ref{eq:unitarity_approximation_bound2}), with the integration performed over $4 m_\pi^2 < s < 2 \, \text{GeV}^2$, whereas the lower part corresponds to the Born approximation, integrated over $2 \, \text{GeV}^2 < s < 4 \, \text{GeV}^2$. Further explanations can be found in the main text.}
\label{tab:CgammagammaiXhadvalues}
\end{table}

To calculate the contributions of intermediate two-body hadronic states to $\big | C_{\gamma \gamma, i}^{\text{had}} (m_h^2) \big |$ in~(\ref{eq:unitarity_approximation_bound1}) and~(\ref{eq:unitarity_approximation_bound2}), one has to evaluate convolutions of $\kappa_i^X(s)$ weighted by $1 / \big | s - m_h^2 - i \epsilon \big |$. Figure~\ref{fig:kappas} displays the functions $\kappa_i^X(s)$ defined in~(\ref{eq:kappaiX}) for the operators $Q_\Gamma$, $Q_\Delta$, and $Q_\theta$ introduced in~(\ref{eq:Qi}), with $X = \pi^+ \pi^-$, $\pi^0 \pi^0$, $K^+ K^-$, and $K_S K_S$. For $4 m_\pi^2 < s < 2 \, \text{GeV}^2$, the shown charged pion and kaon results are obtained from the full $S$-wave cross sections, while for $s > 2 \, \text{GeV}^2$ the Born approximation~(\ref{eq:S-wave_Born}) is employed. Results for the $\pi^0 \eta$ channel are omitted due to the strong form-factor suppression arising from~(\ref{eq:epsilon}) and~(\ref{eq:Fpieta}). From the plots, it is evident that whereas the pion cases exhibit almost smooth transitions between the full and Born results, the kaon cases display pronounced discontinuities. This feature is readily understood from~Figure~\ref{fig:xsectionscomparison}.

\newpage

The behavior of the functions $\kappa_i^X(s)$ shown in~Figure~\ref{fig:kappas} suggests that a conservative estimate of the contributions $\big | C_{\gamma \gamma, i}^{\text{had}} (m_h^2) \big |$ can be obtained as follows. First, instead of performing the integration in~(\ref{eq:unitarity_approximation_bound2}) up to $s = \infty$, we restrict it to $s < 4 \, \text{GeV}^2$. This cut-off is motivated because the contribution of the charm quark should be treatable accurately in perturbation theory. Second, we split the integration into two regions: $4 m_\pi^2 < s < 2 \, \text{GeV}^2$ and $2 \, \text{GeV}^2 < s < 4 \, \text{GeV}^2$. In~the first integration region, the full $S$-wave cross section in~(\ref{eq:unitarity_approximation_bound2}) is used, while the Born approximation~(\ref{eq:S-wave_Born}) is applied in the second region, independently of whether the mesons forming $X$ carry electric charge. Figure~\ref{fig:kappas} shows that, in general, this procedure tends to significantly overestimate the contributions to $\big| C_{\gamma \gamma, i}^{\text{had}} (m_h^2) \big|$ arising from the integration region $2 \, \text{GeV}^2 < s < 4 \, \text{GeV}^2$. Consequently, to obtain a conservative estimate of the hadronic contributions to the $h \to \gamma \gamma$ decay width, this procedure provides a sufficiently accurate approximation. Applying this method, we obtain the numerical values of the magnitudes $\big | C_{\gamma \gamma, i}^{\text{had}} (m_h^2) \big |$ listed in~Table~\ref{tab:CgammagammaiXhadvalues}. Several observations can be made. First, the numbers in the table once again demonstrate that, for each channel $X$, the contributions from~(\ref{eq:unitarity_approximation_bound1}) and~(\ref{eq:unitarity_approximation_bound2}) exhibit a clear hierarchical pattern: the operator $Q_\theta$ yields the largest contribution, followed by $Q_\Delta$, and finally $Q_\Gamma$. Second, the contribution of the $\pi^0 \eta$ channel is negligible, which follows directly from~(\ref{eq:epsilon}) and~(\ref{eq:Fpieta}) and reflects the fact that the $\eta$ meson contains only a small $\pi^0$ admixture. Third, the total contribution from the second integration region is always smaller than that from the first, with the ratios of the second to the first region being approximately $0.07$, $0.53$, and $0.56$ for the operators $Q_\Gamma$, $Q_\Delta$, and $Q_\theta$, respectively. This indicates that summing the results from the two regions provides a conservative bound on the total contribution to $\big | C_{\gamma \gamma, i}^{\text{had}} (m_h^2) \big |$ as defined in~(\ref{eq:unitarity_approximation_bound1}) and~(\ref{eq:unitarity_approximation_bound2}).

Plugging the numerical values of $\big | C_{\gamma \gamma, i}^{\text{had}} (m_h^2) \big |$ from Table~\ref{tab:CgammagammaiXhadvalues} into~(\ref{eq:Cgammagammahad}) and making use of~(\ref{eq:Cgammagamma}) and~(\ref{eq:delta_q}), we obtain
\beq \label{eq:Cgammagammahad_deltahad}
\big | C_{\gamma \gamma}^{\text{had}} \big | \simeq 4.1 \cdot 10^{-8} \,, \qquad \big | \delta_{\text{had}} \big | \simeq 0.0044\% \,, 
\eeq
which, as argued above, should be regarded as conservative estimates of the magnitudes of the Wilson coefficient encoding the hadronic effects in the $h \to \gamma \gamma$ decay width and the corresponding relative shift. We note that numerically $\big | C_{\gamma \gamma}^{\text{had}} \big | \simeq 2 \hspace{0.5mm} \big | C_{\gamma \gamma}^{s} \big |$ and $\big | \delta_{\text{had}} / \delta_s \big | \simeq 2$, with the Wilson coefficient $C_{\gamma \gamma}^{s}$ given in~(\ref{eq:bottom_charm_lightquarks}). Accordingly, in our approach, the hadronic effects are suppressed quadratically by the strange-quark mass, as naively expected. Finally, we add that the numerical value of $\big | C_{\gamma \gamma}^{\text{had}} \big |$ quoted in~(\ref{eq:Cgammagammahad_deltahad}) corresponds to a perturbative contribution from the strange quark~(\ref{eq:bottom_charm_lightquarks}) when $m_s \simeq 130 \, \text{MeV}$ is used. This value of the strange-quark mass corresponds to the $\overline{\text{MS}}$ mass $m_s (1 \, \text{GeV})$, which is approximately a factor of 1.35 larger~\cite{ParticleDataGroup:2024cfk} than $m_s (2 \, \text{GeV}) = 93.5 \, \text{MeV}$, as used in the numerical analysis performed~in~Section~\ref{sec:preliminaries}.

\section{Conclusions} 
\label{sec:conclusions}

Our analysis demonstrated that non-perturbative hadronic contributions to the $h \to \gamma \gamma$ decay width are extremely small, at the level of $0.004\%$, in agreement with the naive expectation that the decay is overwhelmingly dominated by perturbative $W$-boson and top-quark loops. Phenomenologically, this result is important since it implies that the $h \to \gamma \gamma$ decay channel remains one of the cleanest for precision Higgs studies, and the negligible hadronic uncertainties guarantee that the SM prediction is reliable at the $1.7\%$ level~\cite{Davies:2021zbx}, with potential improvements attainable through perturbative calculations of NNLO~EW and mixed QCD-EW corrections~\cite{Sang:2025jfl}.

Within a dispersive framework that respects the symmetries of low-energy QCD, analyticity, and unitarity, we expressed the Wilson coefficient governing hadronic effects as a convolution of light-quark scalar or energy-momentum tensor form factors of the intermediate hadronic states $X$, weighted by the $S$-wave cross sections for $X \to \gamma \gamma$. We~systematically included the most relevant hadronic two-body intermediate states, $X = \pi^+ \pi^-$, $\pi^0 \pi^0$, $\pi^0 \eta$, $K^+ K^-$, and $K_S K_S$, and quantified their contributions numerically. Several~approximations were made in the calculation. However, all are conservative, so our final results~(\ref{eq:Cgammagammahad_deltahad}) remain robust. Numerically, we find that $\big | C_{\gamma \gamma}^{\text{had}} \big | \simeq \alpha/(12 \pi) \hspace{0.5mm} m_s^2/m_h^2 \hspace{0.25mm} \ln^2 \big (m_s^2/m_h^2 \big)$ when $m_s \simeq 130 \, \text{MeV}$ is used, reflecting the expected quadratic mass suppression of light-quark contributions. Our~results differ from those of~\cite{Knecht:2025nyo}, where the hadronic contributions to $h \to \gamma \gamma$ were found to scale linearly with the light-quark masses at the amplitude level, rather than showing the quadratic suppression obtained here. The origin of the discrepancy lies in the treatment of the three-point function for $h \to \gamma \gamma$. Within the lowest-meson dominance approximation used in~\cite{Knecht:2025nyo}, it asymptotically approaches a non-vanishing constant scaling as $\Lambda_{\rm QCD}/v$ in the limit $m_h^2 \to \infty$ for on-shell photons, whereas the dispersive framework yields instead a quadratic suppression with the Higgs-boson mass. We add that, under the assumption of lowest-meson dominance, the hadronic contributions to the $h \to \gamma \gamma$ decay width can also be estimated via $h \to f_0 (500), f_0 (980) \to \gamma \gamma$. Unlike the channels $h \to \pi^0, \eta, \eta^\prime \to \gamma \gamma$ considered in the recent work~\cite{Hernandez-Juarez:2025ees}, the $f_0(500)$ and $f_0 (980)$ resonances provide a non-vanishing contribution to the $h \to \gamma \gamma$ decay width. Details are given in Appendix~\ref{app:LMD}.

Looking forward, several directions could improve and extend this work. One immediate improvement would be to relax the made approximations by including the strong phases appearing in~(\ref{eq:unitarity_approximation}) and using the full $S$-wave cross sections over the entire integration region $4 m_\pi^2 < s < 4 \, \text{GeV}^2$. Additionally, incorporating higher-multiplicity intermediate hadronic states, while expected to be strongly suppressed, could provide a more complete account of hadronic effects and test the convergence of the dispersive expansion. The line of reasoning presented here for $h \to \gamma \gamma$ can, with appropriate modifications, also be applied to compute the hadronic contributions to $h \to \gamma Z$, $h \to gg$, and $gg \to h$. On general grounds, we expect that non-perturbative effects in these cases are safely negligible, remaining at the level of the corresponding perturbative light-quark corrections, which are suppressed quadratically by the quark masses. In~Appendix~\ref{app:ggF}, we substantiate this statement with a simple estimate of the hadronic contributions to Higgs production via ggF. Godspeed to any motivated young particle phenomenologist pursuing the aforementioned improvements and extensions!

\acknowledgments{I would like to thank Marius~Wiesemann for bringing the work~\cite{Knecht:2025nyo} to my attention, and Giulia~Zanderighi for her useful comments on a near-final version of the manuscript, which were helpful in improving it.}

\begin{appendix}

\section{Lowest-meson dominance}
\label{app:LMD}

\begin{figure}[t!]
\centering
\includegraphics[width=0.35\textwidth]{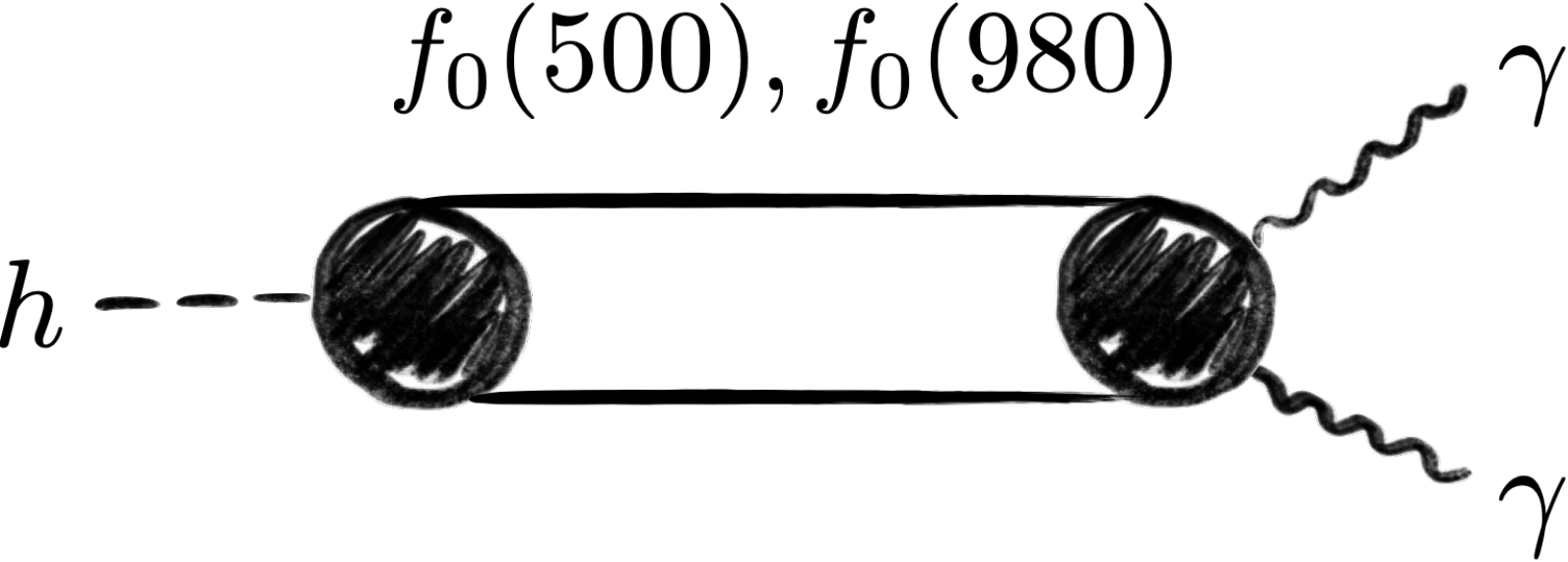}
\vspace{2mm} 
\caption{Graphical representation of the contributions of the $f_0 (500)$ and $f_0 (980)$ resonances to the $h \to \gamma \gamma$ decay. The Higgs mixes with the scalar QCD resonances via the off-diagonal terms $\delta m_{h f_0 (500)}^2$ and $\delta m_{h f_0 (980)}^2$ in the squared mass matrix~(\ref{eq:massmixing}), and the $f_0(500)$ and $f_0 (980)$~mesons subsequently decay to the diphoton final state. \label{fig:lmd}}
\end{figure}

In this appendix, we examine an alternative approach --- albeit a less precise and complete one --- for estimating the magnitude of the hadronic contributions to the $h \to \gamma \gamma$ decay width. This method relies on the quantum mechanical principle that states sharing identical quantum numbers can mix. In our case, this implies that the Higgs boson, which has spin~0 and positive parity, can mix with QCD resonances possessing the same quantum numbers --- namely, the $f_0 (500)$, $f_0 (980)$, and so on. Figure~\ref{fig:lmd} illustrates the contributions of the $f_0 (500)$ and $f_0 (980)$ resonances to the $h \to \gamma \gamma$ decay.

The phenomenon described above can be analyzed using the mass-mixing formalism~\cite{Coleman:1964ac}. Considering a simplified scenario with only two states, the Higgs boson and a scalar state~$S$, the relevant $2 \times 2$ squared mass matrix takes the form
\beq \label{eq:massmixing}
M_{hS}^2 = \begin{pmatrix} m_h^2 - i \hspace{0.125mm} m_h \hspace{0.125mm} \Gamma_h & \delta m_{hS}^2 \\[2mm]
\delta m_{hS}^2 & m_S^2 - i \hspace{0.125mm} m_S \hspace{0.125mm} \Gamma_S \end{pmatrix} \,, 
\eeq
where $m_h$ ($m_S$) and $\Gamma_h$ ($\Gamma_S$) denote the mass and total decay width of the Higgs (scalar) state, respectively, and the off-diagonal term $\delta m_{hS}^2$ encodes the mixing effects. The physical masses and fields are determined from the eigenvalues and eigenvectors of the squared mass matrix~(\ref{eq:massmixing}). In the limit $m_h \gg \Gamma_h, m_S, \Gamma_S, \delta m_{hS}$, which is appropriate and sufficient for our case, the physical Higgs field acquires a small admixture of the bare scalar field $S$, proportional to:
\beq \label{eq:mixingangle}
\sin \phi_{hS} \simeq \frac{\delta m_{hS}^2}{m_h^2} \,. 
\eeq

If the scalar $S$ has a non-vanishing partial decay width to diphotons, the Higgs boson acquires an additional small contribution to its own $h \to \gamma \gamma$ decay width, proportional to~$\Gamma\left(S \to \gamma \gamma\right)$. Defining a Wilson coefficient $C_{\gamma \gamma}^{hS}$, based on~(\ref{eq:Lagrangian}) and~(\ref{eq:width}), which encodes this additional contribution, one can write
\beq \label{eq:CgammagamahS}
\big | C_{\gamma \gamma}^{hS} \big | \simeq \frac{2 \pi v^2}{m_h^3 \hspace{0.25mm} \big | C_{\gamma \gamma} \big |} \hspace{0.125mm} \frac{\delta m_{hS}^2}{m_h^2} \hspace{0.75mm} \Gamma \left ( S \to \gamma \gamma \right ) \,, 
\eeq
where~(\ref{eq:mixingangle}) has been used. 

We now apply the above formalism to the $f_0 (500)$ and $f_0 (980)$ resonances. To estimate the off-diagonal term in~(\ref{eq:massmixing}), it is necessary to consider their internal structure. Various models have been proposed for these light scalar mesons, including conventional quark-antiquark states, tetraquarks, or meson-meson bound states. In practice, the physical resonances may be superpositions of these components, and theoretical approaches are typically used to determine the dominant structure. In the case of the $f_0 (500)$ state, it is often assumed to be a dynamically generated resonance arising from $\pi \pi$ interactions, consistent with the fact that $f_0 (500) \to \pi \pi$ is its dominant (or essentially only) decay mode~\cite{ParticleDataGroup:2024cfk}. For~the $f_0 (980)$ resonance, there is evidence that it is predominantly a $K \bar K$~bound state, since under this assumption the predicted decay widths for $f_0 (980) \to \pi \pi$ and $f_0 (980) \to \gamma \gamma$ are in good agreement with available experimental low-energy data and with results from other theoretical approaches~\cite{ParticleDataGroup:2024cfk}. In the following, we assume that the $f_0 (500)$ and $f_0 (980)$ resonances are pure $\pi\pi$ and $K \bar K$ molecules, respectively. Now, taking into account that the operator $Q_\theta$~in~(\ref{eq:Qi}) has by far the largest form factors $F_i^X(s)$, we estimate
\begin{gather} 
\delta m_{h f_0 (500)}^2 \simeq \frac{2}{9} \hspace{0.5mm} \big | F_\theta^\pi \big ( (500 \, \text{MeV})^2 \big ) \big | \simeq 0.1 \, \text{GeV}^2 \,, \label{eq:dm500} \\[2mm]
\delta m_{h f_0 (980)}^2 \simeq \frac{2}{9} \hspace{0.5mm} \big | F_\theta^K \big ( ( 980 \, \text{MeV})^2 \big ) \big | \simeq 1 \, \text{GeV}^2 \,, \label{eq:dm980}
\end{gather} 
where the factors of $2/9$ stem from~(\ref{eq:Lagrangian}) and the choices of $s$ in the form factors $F_\theta^X (s)$ correspond to on-shell $f_0(500)$ and $f_0 (980)$ resonances.

Besides~(\ref{eq:dm500}) and~(\ref{eq:dm980}), we also need the corresponding partial decay widths to diphotons in order to determine the Wilson coefficients $C_{\gamma \gamma}^{h f_0(500)}$ and $C_{\gamma \gamma}^{h f_0 (980)}$. The relevant partial decay widths are given by
\beq \label{eq:GammaSgammagammas}
\Gamma \left( f_0(500) \to \gamma\gamma \right) \simeq 2 \, \text{keV} \,, \qquad
\Gamma \left( f_0 (980) \to \gamma\gamma \right) = \left( 0.29^{+0.11}_{-0.06} \right) \, \text{keV} \,.
\eeq
Here, the first value represents an approximate average of the individual results quoted in the articles~\cite{Bernabeu:2008wt,Oller:2008kf,Hoferichter:2011wk,Moussallam:2011zg,Dai:2014zta,Danilkin:2020pak}, while the second value is taken directly from the latest PDG~review~\cite{ParticleDataGroup:2024cfk}.

Inserting~(\ref{eq:dm500}), (\ref{eq:dm980}), and (\ref{eq:GammaSgammagammas}) into~(\ref{eq:CgammagamahS}), and using the input parameters summarized in~Table~\ref{tab:input_parameters}, we obtain
\beq \label{eq:CgammagammahSs}
\big | C_{\gamma \gamma}^{h f_0 (500)} \big | \simeq 1.3 \cdot 10^{-9} \,, \qquad \big | C_{\gamma \gamma}^{h f_0 (980)} \big | \simeq 1.9 \cdot 10^{-9} \,.
\eeq
Note that our estimate of the combined contribution of the $f_0 (500)$ and $f_0 (980)$ states amounts to only about $10\%$ of $\big| C_{\gamma \gamma}^{\text{had}} \big|$ given in~(\ref{eq:Cgammagammahad_deltahad}), which represents our best yet conservative estimate of the full hadronic effects in the $h \to \gamma \gamma$ decay width. We consider it a useful consistency check that this estimate, based on the assumption of lowest-meson dominance, falls short of the full result. As noted in~Section~\ref{sec:phenomenology}, this behavior is expected, since both the form factors $F_i^X(s)$ and the $S$-wave cross sections $\sigma_{\text{$S$-wave}} \left (X \to \gamma \gamma \right )$ are significantly influenced by higher-order processes, such as charged meson loops or rescattering. These~effects are only partially captured in the above lowest-meson dominance estimate, which relies on the simple tree-level exchange of the two lightest QCD scalar resonances. In this context, it is also worth noting that replacing~(\ref{eq:dm500}) with~(\ref{eq:dm980}), the combined contribution of the~$f_0(500)$ and $f_0(980)$ resonances accounts for nearly $40\%$ of the full dispersion-relation based estimate given in~(\ref{eq:Cgammagammahad_deltahad}).

The~estimates in this appendix furthermore shows that~\cite{Hernandez-Juarez:2025ees} would have obtained a non-vanishing hadronic contribution to the $h \to \gamma \gamma$ decay width from light quarks if they had considered the $f_0(500)$ and $f_0 (980)$ resonances rather than the $\pi^0$, $\eta$, and $\eta^\prime$ mesons, which cannot mix with the Higgs boson due to their pseudoscalar nature and therefore render no~effect. An estimate of hadronic effects in the $h \to \gamma \gamma$ decay width using a lowest-meson dominance model has also been presented in~\cite{Knecht:2025nyo}. In contrast to our approach, which estimates $\delta m_{hS}^2$ and $\Gamma \left ( S \to \gamma \gamma \right )$ by incorporating both theoretical and experimental input, the work~\cite{Knecht:2025nyo} directly evaluates the three-point function for $h \to \gamma \gamma$ within the lowest-meson dominance approximation, in the limit of large numbers of colors in QCD, $N_c \to \infty$. Due~to the technical differences between the two approaches, a deeper comparison lies beyond the scope of this article.

\section{Hadronic effects in $\bm{gg \to h}$}
\label{app:ggF}

In the following, we outline a simple, though approximate, approach for estimating the magnitude of hadronic contributions to Higgs production via ggF. Our method is based on the assumption that the $f_0 (980)$ resonance is a spin-$0$, parity-even strangeonium state, that~is, a hypothetical bound state of a strange quark and a strange antiquark. Although the~PDG~review~\cite{ParticleDataGroup:2024cfk} does not explicitly list the $f_0 (980)$ as a strangeonium state, recent elliptic anisotropy measurements in proton-lead collisions by the CMS collaboration~\cite{CMS:2023rev} provide evidence that the $f_0 (980)$ resonance behaves as a conventional meson. Let us therefore pursue this idea and examine where it leads.

Due to their relative simplicity, calculations of partial decay widths of quarkonium states into light hadrons, photons, and lepton pairs were among the earliest applications of perturbative QCD. While a rigorous QCD analysis of inclusive annihilation decay widths of quarkonium states can be performed within the effective field theory framework of non-relativistic QCD~(NRQCD)~\cite{Bodwin:1994jh}, for our purposes a LO treatment of $P$-wave annihilation of $s \bar{s}$ bound states is sufficient. The corresponding decay rate of the $f_0 (980)$ resonance into two gluons can be written as~\cite{Barbieri:1975am,Novikov:1977dq}
\beq \label{eq:Gammagg}
\Gamma \left ( f_0 (980) \to gg \right ) \simeq \frac{96 \hspace{0.25mm} \alpha_s^2}{m_{f_0 (980)}^4} \, \big | R^\prime_{f_0 (980)} (0) \big |^2 \,, 
\eeq
where $R^\prime_{f_0 (980)}(0)$ denotes the derivative of the radial wave function of the $f_0 (980)$ meson at the origin. The derivative of the radial wave function of the $f_0 (980)$ meson at the origin also determines the strength of the mixing between the Higgs and the $f_0 (980)$ state. This~mixing can be computed within the NRQCD framework, and to zeroth order in the strong-coupling constant and in the typical velocity of the bound-state quarks, one recovers the results of non-relativistic potential models (see, for instance,~\cite{Drees:1989du}). Expressed as a mixing angle, one can write
\beq \label{eq:dm2}
\sin \varphi_{h f_0 (980)} \simeq \sqrt{\frac{27}{\pi} \, m_{f_0 (980)}} \; \frac{\big | R^\prime_{f_0 (980)} (0) \big |}{v \hspace{0.125mm} m_h^2} \,.
\eeq

In order to calculate the $f_0 (980)$ contribution to Higgs production via ggF, we also need the gluon-gluon luminosity, defined as
\beq \label{eq:gluonlum}
f\hspace{-1.3mm}f_{gg} \left (\tau, \mu_F \right ) \equiv \int_{\tau}^{1} \! \frac{dx}{x} \, f_{g/p} \left (x, \mu_F \right ) \, f_{g/p} \left ( \frac{\tau}{x} , \mu_F \right ) \,,
\eeq
where $f_{g/p}(x, \mu_F)$ denotes the universal, non-perturbative parton distribution function that describes the probability of finding a gluon~($g$) inside the proton~($p$) carrying a longitudinal momentum fraction $x$, and $\mu_F$ is the factorization scale. Using~(\ref{eq:Gammagg}), (\ref{eq:dm2}), and~(\ref{eq:gluonlum}), the $f_0 (980)$ contribution to the $gg \to h$ production cross section can then be expressed~as
\beq \label{eq:sigmaf0980}
\begin{split}
\sigma_{f_0 (980)} \left ( gg \to h \right ) & \simeq f\hspace{-1.3mm}f_{gg} \left (\frac{m_h^2}{S}, m_h^2 \right ) \, \frac{\pi^2}{8 m_{f_0 (980)} \hspace{0.25mm} S} \, \Gamma \left ( f_0 (980) \to gg \right ) \hspace{0.25mm} \sin \varphi_{h f_0 (980)} \\[2mm] 
& \simeq f\hspace{-1.3mm}f_{gg} \left (\frac{m_h^2}{S}, m_h^2 \right ) \, \frac{3}{32 \hspace{0.25mm} \alpha_s} \, \frac{1}{v \hspace{0.125mm} m_h^2 \hspace{0.25mm} S} \, \sqrt{\frac{\pi^3 \hspace{0.5mm} m_{f_0 (980)}^3 \hspace{0.25mm} \Gamma^3 \left ( f_0 (980) \to gg \right )}{2}} \,. 
\end{split}
\eeq
Here, $S$ denotes the center-of-mass energy of the proton-proton collider under consideration, and, in order to arrive at the final result, we have eliminated $\big| R^\prime_{f_0 (980)}(0) \big|$ in favor of $\Gamma\!\left( f_0 (980) \to gg \right)$ by employing~(\ref{eq:Gammagg}).

From~(\ref{eq:sigmaf0980}), it is evident that the size of the $f_0 (980)$ contribution to the $gg \to h$ production cross section depends quite strongly on the assumed $f_0 (980) \to gg$ decay width. Using NRQCD, one can relate the digluon decay width of strangeonium to its diphoton decay width. At LO, one finds~\cite{Barbieri:1975am,Novikov:1977dq}
\beq \label{eq:Gammaggsmall}
\Gamma \left ( f_0 (980) \to gg \right ) \simeq 18 \left ( \frac{\alpha_s}{\alpha} \right )^2 \, \Gamma \left ( f_0 (980) \to \gamma \gamma \right ) \simeq 25 \, \text{MeV} \,, 
\eeq
where the final numerical value corresponds to $\alpha_s \simeq \alpha_s (1 \, \text{GeV}) \simeq 0.5$, the value of the fine-structure constant $\alpha$ from Table~\ref{tab:input_parameters}, and the decay width $\Gamma \left( f_0 (980) \to \gamma \gamma \right)$ quoted in~(\ref{eq:GammaSgammagammas}). Note that, according to the PDG~\cite{ParticleDataGroup:2024cfk}, the result in~(\ref{eq:Gammaggsmall}) corresponds to roughly $25\%$ of the maximum total width, $\Gamma_{f_0 (980)} = 100 \, \text{MeV}$, of the $f_0 (980)$ resonance. A~straightforward upper bound on the $f_0 (980)$ digluon decay width is therefore given by:
\beq \label{eq:Gammagglarge}
\Gamma \left ( f_0 (980) \to gg \right ) \simeq \Gamma_{f_0 (980)} \simeq 100 \, \text{MeV} \,. 
\eeq

For $\sqrt{S} = 13.6 \, \text{TeV}$, $m_{f_0 (980)} \simeq 980 \, \text{MeV}$, the value of $\alpha_s$ quoted above, and the input parameters in~Table~\ref{tab:input_parameters}, the two values in~(\ref{eq:Gammaggsmall}) and~(\ref{eq:Gammagglarge}) lead to 
\beq \label{eq:sigmaf0980s}
\sigma_{f_0 (980)} \left ( gg \to h \right ) \simeq 6.3 \, \text{fb}, \qquad \sigma_{f_0 (980)} \left ( gg \to h \right ) \simeq 50 \, \text{fb} \,,
\eeq
where we have used a gluon-gluon luminosity of $4.1 \cdot 10^6$. The obtained $f_0 (980)$ contributions correspond to roughly $0.01\%$ and $0.1\%$, respectively, of the {\selectlanguage{greek}πρὸς καιρόν}~(pros kairon)~SM~ggF~Higgs production cross section quoted in~\cite{Karlberg:2024zxx}:
\beq \label{eq:dabest}
\sigma \left ( gg \to h \right ) = 52.09 \left ( 1 \pm 5.0\% \right ) \text{pb} \,.
\eeq
Compared to the total uncertainty in~(\ref{eq:dabest}), the possible hadronic contributions to $gg \to h$, estimated in~(\ref{eq:sigmaf0980s}), are clearly negligible. We also note that, for $m_s \simeq 130 \, \text{MeV}$, the perturbative strange-quark contributions yield a relative correction of approximately $0.06\%$ at the one-loop level, corresponding to a cross-section contribution of about $16 \, \text{fb}$. This~value falls neatly between the two estimates given in~(\ref{eq:sigmaf0980s}), indicating that hadronic effects in $gg \to h$ display the expected quadratic mass suppression characteristic of light-quark contributions. We note, in conclusion, that the arguments presented above also imply that the $f_0 (980)$ contributions to the Higgs digluon decay width, $\Gamma_{f_0 (980)} \left ( h \to gg \right )$, amount to approximately 
$0.0002\%$ and $0.001\%$, respectively, of the SM prediction for $\Gamma \left ( h \to gg \right )$, with the first and second numbers corresponding to (\ref{eq:Gammaggsmall}) and (\ref{eq:Gammagglarge}).

\end{appendix}

%\bibliographystyle{apsrev4-1}
%\bibliography{main}

%\end{document}


\end{document}